\documentclass[runningheads]{llncs}
\usepackage[T1]{fontenc}
\usepackage{tabularx}
\usepackage{graphicx}
\usepackage[inline]{enumitem}
\usepackage{booktabs} % For formal tables
\usepackage{multirow}
\usepackage{array}
\usepackage{pifont}
\usepackage{xspace}
\usepackage{subcaption}
\usepackage{makecell}
\usepackage{CJKutf8}
\usepackage{tcolorbox}
\usepackage{acronym}
\usepackage{amsmath}
\usepackage{algorithm}
\usepackage[noend]{algpseudocode}
\usepackage[colorlinks=true,
            linkcolor=blue,
            citecolor=blue,
            filecolor=blue,
            urlcolor=blue]{hyperref}
\acrodef{QG}{query generation}
\acrodef{DSI}{differentiable search index}
\acrodef{GLEN}{generative retrieval via lexical index learning}
\acrodef{GR}{generative retrieval}
\acrodef{LLM}{large language model}
\acrodef{IR}{information retrieval}
\acrodef{NLP}{natural language processing}
\acrodef{docid}{document identifier}

\newcommand{\our}{Few-Shot GR\xspace}
\newcommand{\header}[1]{\vspace*{0.5mm}\noindent\textbf{#1}.}

\begin{document}

\title{Generative Retrieval with Few-shot Indexing}
\titlerunning{Generative Retrieval with Few-shot Indexing}

\author{
Arian Askari\inst{1}\textsuperscript{\textdagger}\orcidID{0000-0003-4712-832X}
\and
Chuan Meng\inst{2}\textsuperscript{\textdagger}\orcidID{0000-0002-1434-7596} \and
Mohammad Aliannejadi\inst{3}\orcidID{0000-0002-9447-4172} \and
Zhaochun Ren\inst{1}\orcidID{0000-0002-9076-6565} \and
Evangelos Kanoulas\inst{3}\orcidID{0000-0002-8312-0694} \and
Suzan Verberne\inst{1}\orcidID{0000-0002-9609-9505
}
}
\authorrunning{Askari et al.}

% First names are abbreviated in the running head. If there are more than two authors, 'et al.' is used.
%
\institute{
Leiden University, Leiden, The Netherlands\\
\email{\{a.askari, z.ren, s.verberne\}@liacs.leidenuniv.nl}
\and
The University of Edinburgh, Edinburgh, United Kingdom\\
\email{chuan.meng@ed.ac.uk}
\and
University of Amsterdam, Amsterdam, The Netherlands\\
\email{\{m.aliannejadi, e.kanoulas\}@uva.nl}
%\\
%[0.5em]
%$^{\dagger}$These authors contributed equally.
%\vspace*{-4mm}
}

\maketitle

\begingroup
\renewcommand{\thefootnote}{} 
\footnotetext{\textsuperscript{\textdagger}These two authors contributed equally.}
\endgroup

\begin{abstract}
Existing \acf{GR} methods rely on \textit{training-based indexing}, which fine-tunes a model to memorise associations between queries and the \acp{docid} of relevant documents.
Training-based indexing suffers from high training costs, under-utilisation of pre-trained knowledge in \acp{LLM}, and limited adaptability to dynamic document corpora.
To address the issues, we propose a \textbf{few-shot} indexing-based \textbf{\ac{GR}} framework (\our).
It has a few-shot indexing process without any training, where we prompt an \ac{LLM} to generate \acp{docid} for all documents in a corpus, ultimately creating a \ac{docid} bank for the entire corpus.
During retrieval, we feed a query to the same \ac{LLM} and constrain it to generate a \ac{docid} within the \ac{docid} bank created during indexing, and then map the generated \ac{docid} back to its corresponding document.
%
%\our does not require any training%, making it more efficient.
%
Moreover, we devise few-shot indexing with \textit{one-to-many mapping} to further enhance \our. 
Experiments show that \our achieves superior performance to state-of-the-art \ac{GR} methods requiring heavy training.

 %  \and Indexing
\end{abstract}

\vspace*{-4mm}
\section{Introduction}
\vspace*{-2.5mm}
\Acf{GR}~\cite{cheng2025descriptive,cai2025exploring,zeng2024planning,zeng2024scalable,kuo2024survey,li2024survey,li2024matching} is a new paradigm in \ac{IR}.
Unlike traditional IR that decouples indexing and retrieval, \ac{GR} unifies both processes into a single model~\cite{tay2022transformer}.
Studies in \ac{GR} typically regard indexing and retrieval as training and inference processes, respectively.
The indexing (training) process typically trains a seq2seq model~\cite{raffel2020exploring} to map queries to the \acp{docid} corresponding to relevant documents, using extensive training data of query--\ac{docid} pairs~\cite{zhuang2022bridging}.
In the retrieval (inference) process, the trained model takes a query text as input and directly generates potentially relevant \acp{docid}.

\header{Limitations}
Existing studies typically rely on \textit{training-based indexing} to memorise the associations between a query and its \ac{docid}.
The nature of training-based indexing has two main limitations:
\begin{enumerate*}[label=(\roman*)]
    \item The approach has a high training overhead~\cite{li2024survey}. 
    Existing studies typically use an \ac{LLM}~\cite{lee2023glen,li2024corpuslm} as the backbone and then fine-tune it with a new learning objective: mapping query text to \acp{docid}.
    Fine-tuning an \ac{LLM} with a new objective demands large-scale query--\ac{docid} pairs, considerable time, and numerous GPUs.
    \item The approach does not make effective use of \acp{LLM}' pre-trained knowledge.
    Because there is a gap between the learning objectives of \acp{LLM} pre-training (text generation) and \ac{GR} fine-tuning (query--\ac{docid} mapping), fine-tuning an \ac{LLM} with \ac{GR}'s objective may cause the \ac{LLM} to forget its pre-trained knowledge~\cite{li2024survey}.
    Little research has explored mainly using \acp{LLM}' pre-trained knowledge for \ac{GR} indexing, without heavy training~\cite{li2024survey}.
    % 
    %\item It is challenging to handle a dynamic corpus. 
    %Training a model to memorise new documents inevitably leads to forgetting old ones~\cite{li2024matching}. 
    %While existing studies propose solutions to mitigate this issue~\cite{mehta2022dsi++,kishore2023incdsi,chen2023continual,guo2024corpusbrain++}, the problem persists due to the inherent nature of training.
\end{enumerate*} 

\header{A new perspective on \ac{GR}}
To address the limitation, we propose a \textbf{few-shot} indexing-based \textbf{\ac{GR}} framework (\our).
Unlike previous \ac{GR} approaches based on training-based indexing, \our has a \textit{few-shot indexing} process, where we index a document corpus without requiring any training.
Specifically, in the few-shot indexing process, \our prompts an \ac{LLM} in a few-shot way to generate a free-text \ac{docid} for each document in a corpus.
This process ultimately produces a \textit{\ac{docid} bank} for all documents in an entire corpus.
%
%Note that unlike the methods proposed by \cite{li2023acid,li2023multiview}, which first generate synthetic \acp{docid} for documents (e.g., by prompting GPT-3.5) and then train another model to learn the mapping from query text to these \acp{docid}, we do not need any training steps.
%
During the retrieval process (inference), the same \ac{LLM} used in few-shot indexing takes a query as input and uses constrained beam search~\cite{de2020autoregressive} to ensure the generated \ac{docid} matches a valid \ac{docid} created during few-shot indexing.

%and comes with various advantages:
%\our fully makes use of an \ac{LLM}'s pre-trained knowledge for indexing, avoids the training overhead of learning the relationship between queries and their \ac{docid}.
%\our can potentially alleviate the challenge of handling dynamic document corpora posed by training-based indexing. This is because \our can easily add or remove \acp{docid} in the \ac{docid} bank created during few-shot indexing, and it does not suffer from the issue of forgetting.
%We, propose few-shot identifier generation without a gap between training and inference and without being limited and fully using pre-training knowledge. 
%While our identifiers are constant, they are not the outcome of training, therefore, our achieved effectiveness is from unseen documents since we don't train the LLM for the task and given new documents added to the corpus, we can still effectively retrieve documents.
%We introduce a novel few-shot method for identifier generation in the context of generative retrieval, moving beyond the traditional dependency on large-scale training data that previous Document-Specific Identifier (DSI) models required. 

%\header{Challenges}
%We argue that \our reduces training overhead and better exploits \acp{LLM}'s pre-trained knowledge.
%
However, the implementation of \our brings one new challenge:
We found that generating only one \ac{docid} per document during few-shot indexing results in limited retrieval quality.
This occurs because a document can be relevant to multiple diverse queries; during retrieval, when the \ac{LLM} is fed with different queries that share the same relevant document, it is hard for the \ac{LLM} to always point to one \ac{docid}.
%During few-shot indexing, we found that when prompting an \ac{LLM} to generate only one \ac{docid} per document,
%During few-shot indexing, we found that when prompting an \ac{LLM} to generate only one \ac{docid} per document, the \ac{LLM} tends to produce the same \ac{docid} for similar documents. 
%This poor differentiation in \acp{docid} ultimately leads to limited retrieval performance.
%\begin{enumerate*}[label=(\roman*)]
    %\item 
    %
    %\item During retrieval, we found that using the open-source implementation of constrained beam search ~\footnote{\url{https://github.com/huggingface/transformers/blob/main/src/transformers/generation/beam_constraints.py}} still decodes invalid \acp{docid}.
    %This occurs because our \ac{LLM} lacks fine-tuning to learn to generate outputs within a specific scope, resulting in more diverse outputs. 
    %Moreover, constrained beam search only controls the validity of the first token generated in the sequence, which, while effective with trained \ac{GR} methods, proves inadequate for \our's needs.
    %found that constrained beam search~\cite{de2020autoregressive} limits \our's retrieval quality. 
    %This limitation arises because constrained beam search only controls the validity of the first token generated in the sequence.
    %\footnote{\url{https://huggingface.co/blog/constrained-beam-search}}
    %the existing constrained beam search based on a prefix tree does not perform well. Challenges
    %current constrained beam search has bank....
    %which is constrained by a prefix tree such that it can only generate document IDs that exist in the corpus. constrained beam search~\cite{}
%\end{enumerate*} 
%\header{Solutions}
We therefore further improve \our to address the challenge.
Unlike most \ac{GR} studies that generate a single \ac{docid} per document, we devise few-shot indexing with \textit{one-to-many mapping}, which enhances few-shot indexing by, for each document, generating multiple \acp{docid}. 
This approach allows a relevant document to be mapped back by multiple various \acp{docid} that are generated in response to different queries during retrieval.

\header{Experiments}
%We equip \our with Llama-3-8B-Instruct~\cite{llama3modelcard} for few-shot indexing and retrieval.
We equip \our with \acp{LLM} for few-shot indexing and retrieval.
Experiments on Natural Questions (NQ)~\cite{kwiatkowski2019natural} and MS MARCO show that \our outperforms or performs comparably to state-of-the-art \ac{GR} methods~\cite{lee2023glen,sun2024learning}.
Moreover, our analyses reveal that two critical factors contribute to the success of \our: 
conducting one-to-many mapping during few-shot indexing, and selecting an effective \ac{LLM}.
Finally, we demonstrate that few-shot indexing is significantly more efficient than training-based indexing.
%, and \our has similar query latency compared to previous \ac{GR} methods

%we analyze how the number of generated \acp{docid} per document during few-shot indexing affects retrieval performance. 
%Our findings indicate that generating 15 \acp{docid} per document achieves optimal performance, and that our generated \acp{docid} are highly diverse.
%
%Moreover, we demonstrate that our proposed custom beam search results in higher retrieval quality than constrained beam search widely used in the \ac{GR} literature.
%
%Besides, we analyze the choice of \acp{LLM} on \our's performance, and we found that \our equipped with other open-source \acp{LLM} still achieves effective retrieval performance.
%
%Finally, we show that \our has lower indexing and inference latency compared to other \ac{GR} methods based on training-based indexing.

%To analyze that, we add ~7k documents from MSMARCO, and then generate identifiers for them and then evaluate the effectiveness on them a

%\header{Reproducibility}

Our main contributions are as follows:
\begin{itemize}[leftmargin=*,nosep]
\item We propose \our, a novel \ac{GR} framework, which conducts \ac{GR} indexing solely with prompting an \ac{LLM} without requiring any training.
%a few-shot manner, fully making use of an LLM’s pre-trained knowledge, avoiding the need for any training steps, and potentially handling well dynamic document corpora.

\item We devise few-shot indexing with one-to-many mapping to further enhance \our's performance.

\item Experiments show that \our achieves superior performance to state-of-the-art \ac{GR} methods that require heavy training.
\end{itemize}

\begin{figure}[t]
  \centering
  \vspace{-2mm}
  \resizebox{\columnwidth}{!}{%
  \begin{tcolorbox}[
      notitle, colback=gray!5, colframe=black,
      boxrule=0.6pt, arc=1mm,
      left=1mm, right=1mm, top=0.6mm, bottom=0.6mm, boxsep=0.6mm
  ]
    \sffamily\footnotesize
    \begin{minipage}[t]{0.49\columnwidth}
      \textbf{Example1}\\
      \textbf{Query:} Provide list of the olympic games?\\
      \textbf{Identifier:} olympic-games-list\\[1mm]

      \textbf{Example3}\\
      \textbf{Query:} How does photosynthesis work in plants?\\
      \textbf{Identifier:} photosynthesis-plant-process
    \end{minipage}\hfill
    \begin{minipage}[t]{0.49\columnwidth}
      \textbf{Example2}\\
      \textbf{Query:} What is minority interest in accounting?\\
      \textbf{Identifier:} subsidiary-corporation-parent\\[1mm]

      \textbf{Example4}\\
      \textbf{Query:} \{new query\}\\
      \textbf{Identifier:}
    \end{minipage}
  \end{tcolorbox}
  }%
  \vspace{-1.5mm}
  \caption{Prompt used for indexing and retrieval. The three queries in the demonstration examples are sampled from NQ's training set~\cite{kwiatkowski2019natural}, while their corresponding \acp{docid} are annotated by the authors.}
  \label{fig:prompt}
  %\vspace{-1mm}
\end{figure}

%\vspace*{-1.5mm}
\section{Methodology}
\vspace*{-4mm}
\label{sec:model}
%\our has two essential steps:
%\begin{enumerate*}[label=(\roman*)]
%    \item few-shot indexing with one-to-many mapping, and
%    \item retrieval with constrained beam search.
%\end{enumerate*} 
%\subsection{Few-shot indexing with one-to-many mapping}
\header{Few-shot indexing with one-to-many mapping}
Let $C=\{d_1,\cdots,$$d_i,\cdots,d_{|C|}\}$ be a corpus with $|C|$ documents; this step aims to use an \ac{LLM} to generate $n$ distinct free-text \acp{docid} $\{id_1,\cdots,id_j,\cdots, id_n\}$ for each document $d$ in the corpus $C$.
Ultimately, we create a \textit{\ac{docid} bank} $B$ that contains \acp{docid} for all documents ($n$ \acp{docid} for each document) in $C$.

%, formally:
%
%\begin{equation}
% docid_j=\mathrm{LLM}(d_i),
%\label{eq:original_indexing}
%\end{equation}
%where $i=1,\cdots,|C|$ and $j=1,\cdots,n$.

Following the \ac{GR} literature~\cite{zhuang2022bridging,pradeep2023does}, which shows that replacing documents with their corresponding pseudo queries during indexing results in better retrieval quality, we use only pseudo queries for indexing.
Specifically, we first generate $n$ pseudo queries $\{\hat{q}_1,\cdots,\hat{q}_j,\cdots, \hat{q}_n\}$ for a document $d_i$ and only feed the generated pseudo queries to the \ac{LLM} to generate $n$ corresponding \acp{docid} $\{id_1,\cdots,id_j,\cdots, id_n\}$, formally:
\begin{equation}
\begin{split}
\hat{q}_j=&\mathrm{QG}(d_i), \\
id_j=&\mathrm{LLM}(\hat{q}_j),
\end{split}
\label{eq:indexing}
\end{equation}
where $QG$ is a pseudo query generator, $i=1,\cdots,|C|$ and $j=1,\cdots,n$.
As depicted in Figure~\ref{fig:prompt}, we prompt the \ac{LLM} in a few-shot manner.

After few-shot indexing, we deduplicate \acp{docid} in the \textit{\ac{docid} bank} $B$.
%At the end of the few-shot indexing process, we remove redundant \acp{docid} from the \textit{\ac{docid} bank} $B$.
The devised one-to-many mapping technique during few-shot indexing effectively captures diverse relevance signals, addressing limitations faced by prior methods relying on single identifier generation per document.

Table~\ref{tab:case} in the appendix gives an example of 10 distinct \acp{docid} generated by \our for a specific document in NQ320K~\cite{lee2023glen,sun2024learning,tay2022transformer}.

%\subsection{Retrieval with constrained beam search}
\header{Retrieval with constrained beam search}
Given a user query $q$ and the \textit{\ac{docid} bank} $B$ created in the previous stage, this step aims to use the same prompt (see Figure~\ref{fig:prompt}) and the \ac{LLM} (see Equation~\ref{eq:indexing}) from the indexing phase to generate a \ac{docid} $id$, formally:
\begin{equation}
id = \mathrm{LLM}(q),
\label{eq:retrieval}
\end{equation}
Where we use constrained beam search~\cite{de2020autoregressive} to the \ac{LLM}'s decoding, ensuring the generated \ac{docid} $id$ matches a valid \ac{docid} in the \textit{\ac{docid} bank} $B$. 
Finally, we map the matched valid \ac{docid} back to its corresponding document.
Note that the \textit{\ac{docid} bank} $B$ undergoes de-duplication, ensuring that each \ac{docid} uniquely corresponds to a single document.

\vspace*{-3mm}
\section{Experimental setup}
\vspace*{-3mm}
\label{sec:setup}

\header{Datasets}
We evaluate on NQ320K~\cite{lee2023glen,sun2024learning,tay2022transformer} and MS300K~\cite{wang2023novo,mekonnen2025lightweight}; both have widely been used for \ac{GR} evaluation. 
NQ320K is a version of Natural Questions (NQ)~\cite{kwiatkowski2019natural}; NQ320K consists of 320k relevant query--document pairs, 100k documents, and 7,830 test queries.
MS300K is a version of MS MARCO; MS300K contains 300k query–document pairs, 320k documents, and 5,187 test queries.

%Following recent studies~\cite{lee2023glen,sun2024learning}, we fetch and process NQ320K using the script released by \cite{wang2022neural},\footnote{\url{https://github.com/solidsea98/Neural-Corpus-Indexer-NCI}} to ensure our results are comparable with previous work.

\header{Baselines}
We use non-\ac{GR} and \ac{GR} baselines.
Following \cite{lee2023glen}, we use the following non-\ac{GR} baselines: BM25, DPR~\cite{karpukhin2020dense}, SentenceT5~\cite{ni2022sentence}, and GTR-base~\cite{ni2022large}.
We use the following \ac{GR} baselines (training-based indexing):
\begin{enumerate*}[label=(\roman*)]
%\begin{itemize}[leftmargin=*,nosep]
    \item SEAL~\cite{bevilacqua2022autoregressive}  learns to generate n-grams-based \acp{docid} and applies FM-index~\cite{ferragina2000opportunistic}. %to ensure valid \ac{docid} generation.
    \item DSI~\cite{tay2022transformer} learns to generate numeric identifiers.%\cm{@Arian, which identifier type you used for implementing this baseline?} Arian: we report numbers from their paper. For DSI-QG with query generated by InPARS, I have followed their implementation as it is publicly available and the identifer type in it is arbitrary strings.
    \item DSI-QG~\cite{zhuang2022bridging} augments DSI training by using pseudo queries; we replicate DSI-QG using the pseudo query generator provided by the original paper. 
    \item DSI-QG (InPars) uses the pseudo query generator from InPars~\cite{bonifacio2022inpars}. 
    \item TOME~\cite{ren2023tome} learns to generate document URLs.
    \item GLEN~\cite{lee2023glen} learns dynamic lexical \acp{docid}.
    \item GenRET~\cite{sun2024learning} learns to assign numeric \acp{docid} based on an auto-encoding scheme.
    \item NOVO~\cite{wang2023novo} learns interpretable \acp{docid}.
%\end{itemize}
\end{enumerate*}

\begin{table}[t]
\centering
\caption{
Retrieval quality of \our and baselines on NQ320K and MS300K.
DSI-QG (InPars) and \our use the query generator from InPars~\cite{bonifacio2022inpars} to generate pseudo queries.
Methods marked $^{\dagger}$ are our reimplementations; all other results are from the corresponding papers~\cite{wang2023novo,sun2024learning,ren2023tome,wang2023novo}.
The best value in each column is marked in \textbf{bold}, and the second best is \underline{underlined}.
}
\label{tab:result}
\setlength{\tabcolsep}{3.5pt}
\renewcommand{\arraystretch}{1.05}
\resizebox{\columnwidth}{!}{%
\begin{tabular}{l ccc ccc}
\toprule
\multirow{2}{*}{\textbf{Method}} &
\multicolumn{3}{c}{\textbf{NQ320K}} &
\multicolumn{3}{c}{\textbf{MS300K}} \\
\cmidrule(lr){2-4}\cmidrule(lr){5-7}
& Recall@1 & Recall@10 & MRR@100 & Recall@1 & Recall@10 & MRR@10 \\
\midrule
BM25                 & 29.7 & 60.3 & 40.2 & 39.1 & 69.1 & 48.6 \\
DocT5Query           & 38.0 & 69.3 & 48.9 & 46.7  & 76.5 & 56.2 \\
%DPR                  & 50.2 & 77.7 & 59.9 & -- & -- & -- \\
ANCE                 & 50.2 & 78.5 & 60.2 & 45.6 & 75.7 & 55.6 \\
SentenceT5           & 53.6 & 83.0 & 64.1 & 41.8 & 75.4 & 52.8 \\
GTR-base             & 56.0 & 84.4 & 66.2 & -- & -- & -- \\
\midrule
SEAL                 & 59.9 & 81.2 & 67.7 & 25.9 & 68.6 & 40.2 \\
DSI                  & 55.2 & 67.4 & 59.6 & 32.4  & 69.9 & 44.3 \\
NCI                  & 66.4 & 85.7 & 73.6 & 30.1 & 64.3 & 41.7 \\
DSI-QG$^{\dagger}$               & 63.1 & 80.7 & 69.5 & 41.0 & 71.2 & 50.7 \\
DSI-QG (InPars)$^{\dagger}$      & 63.9 & 82.0 & 71.4 & 41.3 & 71.5 & 50.0 \\
TOME                 & 66.6 & --   & --   & -- & -- & -- \\
GLEN                 & 69.1 & 86.0 & 75.4 & -- & -- & -- \\
GenRET               & 68.1 & \underline{88.8} & 75.9 & 47.9 & 79.8 & 58.1 \\
NOVO  & \underline{69.3} & \textbf{89.7} & \underline{76.7} & \underline{49.1} & \underline{80.8} & \textbf{59.2}\\
\midrule
\our   & \textbf{70.1} & 87.6 & \textbf{77.4}  & \textbf{49.6} & \textbf{81.2} & \underline{59.1}  \\
\bottomrule
\end{tabular}}
\end{table}

\header{Evaluation metrics}
In line with recent \ac{GR} work~\cite{wang2023novo,lee2023glen,sun2024learning}, we report Recall@{1,10} on both datasets, plus MRR@100 (NQ320K) and MRR@10 (MS300K).

\header{Implementation details}
We equip \our with llama-3-8B-Instruct for indexing and retrieval.%\footnote{\url{https://huggingface.co/meta-llama/Meta-Llama-3-8B-Instruct}}
We generate 10 \acp{docid} per document during few-shot indexing.
%
%We use a beam size of 200 for inference.~\cm{@Arian, what is the beam size used by others?}
%
We set the maximum and minimum lengths for \ac{docid} generation to 15 and 3 tokens, respectively.
We employ the query generator from InPars~\cite{bonifacio2022inpars} for generating pseudo queries in Equation~\ref{eq:indexing}.
We conduct parameter tuning on the training set of NQ320K or MS300K.
%\cm{@Arian, please add details into parameter tuning.}
%\cm{@Arian, what are the settings of baselines? Please add more details:  e.g., beam size, do we directly use their paper numbers or do we replicate them?}

% We follow the publicly available implementation of DSI-QG for replication, replacing DOCTTTQuery with InPars. @Chuan, feel free to add it if you decide to include the results of DSI-QG with InPars.

%

\vspace*{-3mm}
\section{Result and analysis}
\vspace*{-3.5mm}
\label{sec:result}

\header{Comparison with baselines}
Table~\ref{tab:result} shows the retrieval quality of \our and all baselines on NQ320K and MS300K.
The leading observation is that \our outperforms all baselines across all metrics, except GenRET/NOVO on Recall@10 (NQ320K)/MRR@10 (MS300K).
This shows that our proposed few-shot indexing is highly effective versus training-based indexing.
Notably, while GenRET/NOVO is slightly better on those metrics, it requires large training corpora and heavy model-specific training, which may not be feasible in low-resource settings. 
In contrast, \our achieves strong results using only a small set of examples, making it more practical.

\begin{figure}[t]
    \centering
    \vspace{-1mm} 
    \includegraphics[width=0.65\linewidth]{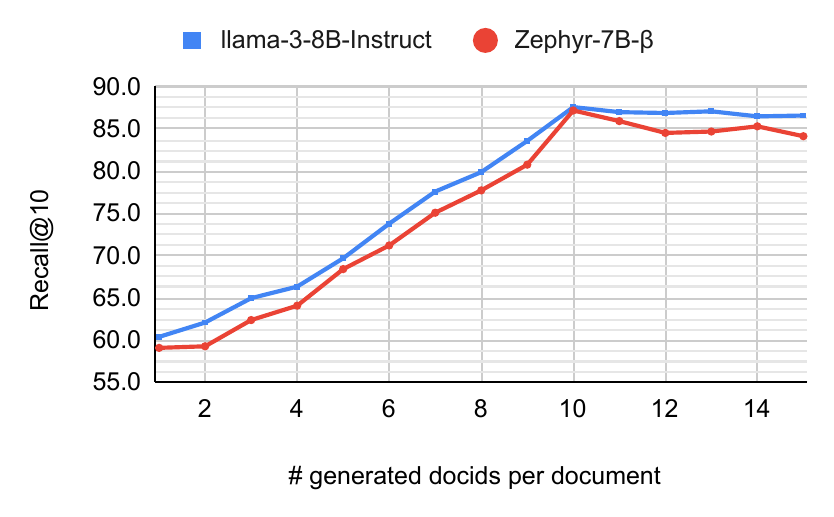}
    \vspace{-2.5mm} 
    \caption{\our's retrieval quality w.r.t.\ \# generated \acp{docid} per document in few-shot indexing on NQ320K.}
    \vspace{-1.5mm} 
    \label{fig:num}
\end{figure}

\header{The impact of \# \acp{docid} generated per document}
%\label{sec:num}
Figure~\ref{fig:num} shows \our's performance w.r.t.\ \# generated \acp{docid} per document during few-shot indexing on NQ320K; we equip \our with llama-3-8B-Instruct or Zephyr-7B-$\beta$~\cite{tunstall2023zephyr}.
%\footnote{\url{https://huggingface.co/HuggingFaceH4/zephyr-7b-beta}}
%
We found that \our's performance improves as it generates more \acp{docid} per document during indexing, reaching saturation when generating 10 \acp{docid}.
E.g., with Llama-3, increasing the number of generated \acp{docid} from 1 to 10 yields a 27.2\% improvement in Recall@10.
It suggests that our devised ``one-to-many mapping'' is key to the success of few-shot indexing.
The trend is similar on MS300K; we report only NQ320K hereafter due to space.
%

%Table~\ref{tab:case} in the appendix gives an example of 10 distinct \acp{docid} generated by \our for a specific document.

%Figure \ref{fig:num} illustrate the relation within the number of generated ids per document and the effectiveness of our proposed method. We can observe that there is a large difference within one id and ten ids to be used and after using more than 10 ids a slight drop on effectiveness is observed.

\begin{table}[t]
\centering
\caption{
Retrieval quality of \our with different \acp{LLM} on NQ320K.
%
%\our always generates 10 \acp{docid} per document during indexing.
%
%The best value in each column is marked in \textbf{bold}.
}
\label{tab:llm} 
%\setlength{\tabcolsep}{1mm}
%\scalebox{0.85}{
    \begin{tabular}{l ccc}
    \toprule
      \multirow{1}{*}{{\textbf{Method}}} & Recall@1     & Recall@10  & MRR@100 \\
    \midrule
    T5-base  &  52.4	& 66.4	& 55.8  \\
    Zephyr-7B-$\beta$  &  69.9 &	87.2 & \textbf{77.8}  \\
    llama-3-8B-Instruct  & \textbf{70.1}  & \textbf{87.6} & 77.4  \\
    \bottomrule
    \end{tabular}
%}
\end{table}

\header{The impact of \acp{LLM} choices}
%\label{sec:llm}
Table~\ref{tab:llm} shows \our's performance using different \acp{LLM} on NQ320K; here we compare T5-base, Zephyr-7B-$\beta$, and llama-3-8B-Instruct.
We found that Llama-3-8B-Instruct performs the best across most metrics, followed by Zephyr-7B-$\beta$. 
However, both markedly outperform T5-base in terms of performance.
It suggests that selecting an effective \ac{LLM} is another critical factor contributing to the success of \our.

\begin{table}[t]
\centering
\caption{
Efficiency of indexing and retrieval for \our and training-based \ac{GR} baselines on NQ320K.
\our uses llama-3-8B-Instruct and generates 10 \acp{docid} per document during few-shot indexing.
%
%hr and ms denote hours and milliseconds, respectively.
}
\label{tab:cost} 
%\setlength{\tabcolsep}{1mm}
%\scalebox{0.85}{
\begin{tabular}{l cc}
\toprule
  \multirow{1}{*}{{\textbf{Method}}} & Indexing (hr)     & Retrieval (ms) \\
\midrule
DSI-QG &  240 &   72  \\
GenRET   & $\approx$16,800  &   72  \\
\midrule
\our  &  37 &   98  \\
\bottomrule
\end{tabular}
%}
\end{table}

\header{Efficiency of indexing and retrieval}
Table~\ref{tab:cost} presents the indexing time and retrieval latency for \our compared to two training-based \ac{GR} methods, DSI-QG~\cite{zhuang2022bridging} and GenRET~\cite{sun2024learning}.
The time cost of indexing is measured in hours (hr) on the training set of NQ320K, while the retrieval query latency is measured in milliseconds (ms) on the test set of NQ320K.
We perform all measurements on a single A100 GPU (80GB) with a batch size of 16, except for the indexing (training) of GenRET. 
We inquired with the authors of GenRET~\cite{sun2024learning} about GenRET's indexing (training) time, and they indicated it took 7 days on 100 A100 GPUs. 
This implies it may take approximately 16,800 hours on a single A100 GPU.
We found that \our is significantly more efficient in indexing than existing \ac{GR} methods.
Also, \our achieves similar retrieval query latency compared to existing \ac{GR} methods.
\vspace*{-3mm}
\section{Conclusions \& Future Work}
\vspace*{-3mm}
We have proposed a new, efficient, and effective \ac{GR} paradigm, \our, featuring a few-shot indexing process that solely relies on prompting an \ac{LLM} to record associations between queries and their \acp{docid}, eliminating the need for any training steps.
%
%In this process, we prompt an \ac{LLM} to generate \acp{docid} for all documents in a corpus, so as to create a \ac{docid} bank for the entire corpus.
%
%Given a query, \our conducts retrieval by promoting the \ac{LLM} used for indexing and constraining it to generate a \ac{docid} within the recorded \ac{docid} created in indexing.
%
We have designed \textit{few-shot indexing with one-to-many mapping} to further enhance \our's indexing.
Experimental results show that \ac{GR} achieves superior performance to training-intensive state-of-the-art \ac{GR} methods.

\header{Suitability for dynamic corpora}
Training-based indexing struggles with dynamic corpora, as training on new documents often causes forgetting of old ones~\cite{li2024matching}.
Although several studies attempt to mitigate this issue~\cite{mehta2022dsi++,kishore2023incdsi,chen2023continual,guo2024corpusbrain++}, it remains inherent to training-based methods.
\our alleviates this challenge by enabling easy addition or removal of \acp{docid} in the few-shot indexing \ac{docid} bank, thus avoiding catastrophic forgetting.
Future work can further explore this direction.

%\header{Limitation}
The datasets used in this paper, NQ320K and MS300K, contain corpora of 100K and 320K documents, respectively.
So it is worthwhile to test whether \our's effectiveness would generalise to a million-document corpus.
Also, it is worth testing \our on other datasets (e.g., BEIR~\cite{thakur2021beir} and conversational search domains~\cite{mo2025conversational,meng2026conversational1,meng2025bridging,meng2023query}).
Finally, exploring automatic retrieval quality prediction for generative retrieval methods is another promising direction~\cite{10.1145/3769733.3769743,meng2025query}.

\clearpage

\bibliographystyle{splncs04}
\bibliography{references}

@article{10.1145/3769733.3769743,
author = {Meng, Chuan and Faggioli, Guglielmo and Aliannejadi, Mohammad and Ferro, Nicola and Mothe, Josiane},
title = {Report on the 2nd Workshop on Query Performance Prediction and its Applications in the Era of Large Language Models ({QPP++} 2025) at {ECIR} 2025},
year = {2025},
issue_date = {June 2025},
publisher = {Association for Computing Machinery},
address = {New York, NY, USA},
volume = {59},
number = {1},
issn = {0163-5840},
pages = {1–8},
numpages = {8}
}

@inproceedings{meng2023query,
  title={Query Performance Prediction: From Ad-hoc to Conversational Search},
  author={Meng, Chuan and Arabzadeh, Negar and Aliannejadi, Mohammad and De Rijke, Maarten},
  booktitle={SIGIR},
  pages={2583--2593},
  year={2023}
}

@inproceedings{meng2026conversational1,
  title={Conversational Search: From Fundamentals to Frontiers in the Age of Agents},
  author={Meng, Chuan and Mo, Fengran and Aliannejadi, Mohammad and Dalton, Jeff and Nie, Jian-Yun},
  booktitle={WWW},
  pages={},
  year={2026}
}

@inproceedings{meng2025bridging,
  title={Bridging the Gap: From Ad-hoc to Proactive Search in Conversations},
  author={Meng, Chuan and Tonolini, Francesco and Mo, Fengran and Aletras, Nikolaos and Yilmaz, Emine and Kazai, Gabriella},
  booktitle={SIGIR},
  pages={64--74},
  year={2025}
}

@inproceedings{mo2025conversational,
  title={Conversational search: From Fundamentals to Frontiers in the {LLM} Era},
  author={Mo, Fengran and Meng, Chuan and Aliannejadi, Mohammad and Nie, Jian-Yun},
  booktitle={SIGIR},
  pages={4094--4097},
  year={2025}
}

@article{meng2025query,
  title={Query Performance Prediction using Relevance Judgments Generated by Large Language Models},
  author={Meng, Chuan and Arabzadeh, Negar and Askari, Arian and Aliannejadi, Mohammad and de Rijke, Maarten},
  journal={TOIS},
  volume={43},
  number={4},
  pages={1--35},
  year={2025},
  publisher={ACM New York, NY}
}

@inproceedings{cheng2025descriptive,
  title={Descriptive and Discriminative Document Identifiers for Generative Retrieval},
  author={Cheng, Jiehan and Dou, Zhicheng and Zhu, Yutao and Li, Xiaoxi},
  booktitle={AAAI},
  volume={39},
  number={11},
  pages={11518--11526},
  year={2025}
}

@inproceedings{mekonnen2025lightweight,
  title={Lightweight and Direct Document Relevance Optimization for Generative Information Retrieval},
  author={Mekonnen, Kidist Amde and Tang, Yubao and de Rijke, Maarten},
  booktitle={SIGIR},
  pages={1327--1338},
  year={2025}
}

@inproceedings{wang2023novo,
  title={NOVO: Learnable and Interpretable Document Identifiers for Model-Based IR},
  author={Wang, Zihan and Zhou, Yujia and Tu, Yiteng and Dou, Zhicheng},
  booktitle={CIKM},
  pages={2656--2665},
  year={2023}
}

@inproceedings{cai2025exploring,
  title={Exploring Training and Inference Scaling Laws in Generative Retrieval},
  author={Cai, Hongru and Li, Yongqi and Yuan, Ruifeng and Wang, Wenjie and Zhang, Zhen and Li, Wenjie and Chua, Tat-Seng},
  booktitle={SIGIR},
  pages={1339--1349},
  year={2025}
}

@inproceedings{thakur2021beir,
  title={BEIR: A Heterogeneous Benchmark for Zero-shot Evaluation of Information Retrieval Models},
  author={Thakur, Nandan and Reimers, Nils and R{\"u}ckl{\'e}, Andreas and Srivastava, Abhishek and Gurevych, Iryna},
  booktitle={Thirty-fifth Conference on Neural Information Processing Systems Datasets and Benchmarks Track (Round 2)},
  year={2021}
}

@inproceedings{bonifacio2022inpars,
author = {Bonifacio, Luiz and Abonizio, Hugo and Fadaee, Marzieh and Nogueira, Rodrigo},
title = {{InPars}: Unsupervised Dataset Generation for Information Retrieval},
year = {2022},
isbn = {9781450387323},
booktitle = {SIGIR},
pages = {2387–2392},
numpages = {6},
}

@inproceedings{ren2023tome,
  title={TOME: A Two-stage Approach for Model-based Retrieval},
  author={Ren, Ruiyang and Zhao, Wayne Xin and Liu, Jing and Wu, Hua and Wen, Ji-Rong and Wang, Haifeng},
  booktitle={ACL},
  pages={6102--6114},
  year={2023}
}

@inproceedings{ni2022large,
  title={Large Dual Encoders Are Generalizable Retrievers},
  author={Ni, Jianmo and Qu, Chen and Lu, Jing and Dai, Zhuyun and Abrego, Gustavo Hernandez and Ma, Ji and Zhao, Vincent and Luan, Yi and Hall, Keith and Chang, Ming-Wei and others},
  booktitle={EMNLP},
  pages={9844--9855},
  year={2022}
}

@inproceedings{ni2022sentence,
  title={Sentence-T5: Scalable Sentence Encoders from Pre-trained Text-to-Text Models},
  author={Ni, Jianmo and Abrego, Gustavo Hernandez and Constant, Noah and Ma, Ji and Hall, Keith and Cer, Daniel and Yang, Yinfei},
  booktitle={Findings of ACL},
  pages={1864--1874},
  year={2022}
}

@inproceedings{karpukhin2020dense,
  title={Dense Passage Retrieval for Open-Domain Question Answering},
  author={Karpukhin, Vladimir and Oguz, Barlas and Min, Sewon and Lewis, Patrick and Wu, Ledell and Edunov, Sergey and Chen, Danqi and Yih, Wen-tau},
  booktitle={EMNLP},
  pages={6769--6781},
  year={2020}
}

@inproceedings{de2020autoregressive,
  title={Autoregressive Entity Retrieval},
  author={De Cao, Nicola and Izacard, Gautier and Riedel, Sebastian and Petroni, Fabio},
  booktitle={ICLR},
  year={2020}
}

@article{li2024corpuslm,
  title={CorpusLM: Towards a Unified Language Model on Corpus for Knowledge-Intensive Tasks},
  author={Li, Xiaoxi and Dou, Zhicheng and Zhou, Yujia and Liu, Fangchao},
  year={2024}
}

@article{guo2024corpusbrain++,
  title={CorpusBrain++: A Continual Generative Pre-Training Framework for Knowledge-Intensive Language Tasks},
  author={Guo, Jiafeng and Zhou, Changjiang and Zhang, Ruqing and Chen, Jiangui and de Rijke, Maarten and Fan, Yixing and Cheng, Xueqi},
  journal={arXiv preprint arXiv:2402.16767},
  year={2024}
}

@inproceedings{chen2023continual,
  title={Continual Learning for Generative Retrieval over Dynamic Corpora},
  author={Chen, Jiangui and Zhang, Ruqing and Guo, Jiafeng and de Rijke, Maarten and Chen, Wei and Fan, Yixing and Cheng, Xueqi},
  booktitle={CIKM},
  pages={306--315},
  year={2023}
}

@inproceedings{kishore2023incdsi,
  title={IncDSI: Incrementally Updatable Document Retrieval},
  author={Kishore, Varsha and Wan, Chao and Lovelace, Justin and Artzi, Yoav and Weinberger, Kilian Q},
  booktitle={International Conference on Machine Learning},
  pages={17122--17134},
  year={2023},
  organization={PMLR}
}

@inproceedings{ferragina2000opportunistic,
  title={Opportunistic Data Structures with Applications},
  author={Ferragina, Paolo and Manzini, Giovanni},
  booktitle={Proceedings 41st annual symposium on foundations of computer science},
  pages={390--398},
  year={2000},
  organization={IEEE}
}

@article{raffel2020exploring,
  title={Exploring the Limits of Transfer Learning with a Unified Text-to-Text Transformer},
  author={Raffel, Colin and Shazeer, Noam and Roberts, Adam and Lee, Katherine and Narang, Sharan and Matena, Michael and Zhou, Yanqi and Li, Wei and Liu, Peter J},
  journal={JMLR},
  volume={21},
  number={140},
  pages={1--67},
  year={2020}
}

@article{kwiatkowski2019natural,
  title={Natural Questions: A Benchmark for Question Answering Research},
  author={Kwiatkowski, Tom and Palomaki, Jennimaria and Redfield, Olivia and Collins, Michael and Parikh, Ankur and Alberti, Chris and Epstein, Danielle and Polosukhin, Illia and Devlin, Jacob and Lee, Kenton and others},
  journal={TACL},
  volume={7},
  pages={453--466},
  year={2019},
  publisher={MIT Press One Rogers Street, Cambridge, MA 02142-1209, USA journals-info~…}
}

@article{mehta2022dsi++,
  title={DSI++: Updating Transformer Memory with New Documents},
  author={Mehta, Sanket Vaibhav and Gupta, Jai and Tay, Yi and Dehghani, Mostafa and Tran, Vinh Q and Rao, Jinfeng and Najork, Marc and Strubell, Emma and Metzler, Donald},
  journal={arXiv preprint arXiv:2212.09744},
  year={2022}
}

@article{pradeep2023does,
  title={How Does Generative Retrieval Scale to Millions of Passages?},
  author={Pradeep, Ronak and Hui, Kai and Gupta, Jai and Lelkes, Adam D and Zhuang, Honglei and Lin, Jimmy and Metzler, Donald and Tran, Vinh Q},
  journal={arXiv preprint arXiv:2305.11841},
  year={2023}
}

@inproceedings{lee2023glen,
  title={GLEN: Generative Retrieval via Lexical Index Learning},
  author={Lee, Sunkyung and Choi, Minjin and Lee, Jongwuk},
  booktitle={EMNLP},
  pages={7693--7704},
  year={2023}
}

@article{zhuang2022bridging,
  title={Bridging the Gap Between Indexing and Retrieval for Differentiable Search Index with Query Generation},
  author={Zhuang, Shengyao and Ren, Houxing and Shou, Linjun and Pei, Jian and Gong, Ming and Zuccon, Guido and Jiang, Daxin},
  journal={arXiv preprint arXiv:2206.10128},
  year={2022}
}

@article{tunstall2023zephyr,
  title={Zephyr: Direct Distillation of LM Alignment},
  author={Tunstall, Lewis and Beeching, Edward and Lambert, Nathan and Rajani, Nazneen and Rasul, Kashif and Belkada, Younes and Huang, Shengyi and von Werra, Leandro and Fourrier, Cl{\'e}mentine and Habib, Nathan and others},
  journal={arXiv preprint arXiv:2310.16944},
  year={2023}
}

@article{li2024survey,
  title={A Survey of Generative Search and Recommendation in the Era of Large Language Models},
  author={Li, Yongqi and Lin, Xinyu and Wang, Wenjie and Feng, Fuli and Pang, Liang and Li, Wenjie and Nie, Liqiang and He, Xiangnan and Chua, Tat-Seng},
  journal={arXiv preprint arXiv:2404.16924},
  year={2024}
}

@article{zeng2024planning,
  title={Planning Ahead in Generative Retrieval: Guiding Autoregressive Generation through Simultaneous Decoding},
  author={Zeng, Hansi and Luo, Chen and Zamani, Hamed},
  journal={arXiv preprint arXiv:2404.14600},
  year={2024}
}

@inproceedings{zeng2024scalable,
  title={Scalable and Effective Generative Information Retrieval},
  author={Zeng, Hansi and Luo, Chen and Jin, Bowen and Sarwar, Sheikh Muhammad and Wei, Tianxin and Zamani, Hamed},
  booktitle={WWW},
  pages={1441--1452},
  year={2024}
}

@article{sun2024learning,
  title={Learning to Tokenize for Generative Retrieval},
  author={Sun, Weiwei and Yan, Lingyong and Chen, Zheng and Wang, Shuaiqiang and Zhu, Haichao and Ren, Pengjie and Chen, Zhumin and Yin, Dawei and Rijke, Maarten and Ren, Zhaochun},
  journal={NeurIPS},
  volume={36},
  year={2024}
}

@article{li2024matching,
  title={From Matching to Generation: A Survey on Generative Information Retrieval},
  author={Li, Xiaoxi and Jin, Jiajie and Zhou, Yujia and Zhang, Yuyao and Zhang, Peitian and Zhu, Yutao and Dou, Zhicheng},
  journal={arXiv preprint arXiv:2404.14851},
  year={2024}
}

@inproceedings{bevilacqua2022autoregressive,
  title={Autoregressive Search Engines: Generating Substrings as Document Identifiers},
  author={Bevilacqua, Michele and Ottaviano, Giuseppe and Lewis, Patrick and Yih, Scott and Riedel, Sebastian and Petroni, Fabio},
  booktitle={NeurIPS},
  volume={35},
  pages={31668--31683},
  year={2022}
}

@inproceedings{tay2022transformer,
  title={Transformer Memory as a Differentiable Search Index},
  author={Tay, Yi and Tran, Vinh and Dehghani, Mostafa and Ni, Jianmo and Bahri, Dara and Mehta, Harsh and Qin, Zhen and Hui, Kai and Zhao, Zhe and Gupta, Jai and others},
  booktitle={NeurIPS},
  volume={35},
  pages={21831--21843},
  year={2022}
}

@article{kuo2024survey,
  title={A Survey of Generative Information Retrieval},
  author={Kuo, Tzu-Lin and Chiu, Tzu-Wei and Lin, Tzung-Sheng and Wu, Sheng-Yang and Huang, Chao-Wei and Chen, Yun-Nung},
  journal={arXiv preprint arXiv:2406.01197},
  year={2024}
}

\clearpage
%\appendix
\section*{Appendix}
\label{sec:appendix}
%\vspace*{-3mm}

\header{Case study of \acp{docid} generated by \our}
Table~\ref{tab:case} gives an example of 10 distinct \acp{docid} generated by \our for a specific document in NQ320K.
It shows that \acp{docid} generated by \our are various.

\begin{table}[!ht]
\centering
\caption{
Pseudo queries and corresponding \acp{docid} generated by \our using Llama-3-8B-Instruct for a document from NQ320K.
}
\label{tab:case} 
\setlength{\tabcolsep}{1mm}
\begin{tabular}{p{0.4\textwidth} p{0.28\textwidth} p{0.28\textwidth}}
\toprule
\textbf{Document text} & \textbf{Pseudo queries} & \textbf{\Acp{docid}} \\
\midrule
\multirow{12}{0.4\textwidth}{In accounting, minority interest (or non-controlling interest) is the portion of a subsidiary corporation's stock that is not owned by the parent corporation. The magnitude of the minority interest in the subsidiary company is generally less than 50\% of outstanding shares, or the corporation would generally cease to be a subsidiary of the parent.
} & What is minority interest in accounting? & minority-interest-accounting  \\ \cmidrule(lr){2-3}
 & What is non-controlling interest in accounting?  & non-controlling-interest-accounting \\ \cmidrule(lr){2-3}
 & How is minority interest defined in accounting?  & minority-interest-definition \\ \cmidrule(lr){2-3}
 & How is minority interest calculated in accounting?  & minority-interest-calculation \\ \cmidrule(lr){2-3}
 & What is the significance of minority interest in accounting?  & minority-interest-significance
 \\ \cmidrule(lr){2-3}
 & How does minority interest affect financial statements in accounting?  & minority-interest-financial-statements \\ \cmidrule(lr){2-3}
 & How is minority interest treated in consolidated financial statements in accounting?  & minority-interest-consolidated-financial-statements \\ \cmidrule(lr){2-3}
 & What is the impact of minority interest on the parent company's earnings per share in accounting?  & minority-interest-impact-eps
 \\ \cmidrule(lr){2-3}
 & How is minority interest reported in financial statements of a subsidiary company in accounting?  & minority-interest-reporting-subsidiary
 \\ \cmidrule(lr){2-3}
 & What is the role of minority interest in a parent corporation's financial statements?  & minority-interest-parent-corporation \\ 
\bottomrule
\end{tabular}
\end{table}

\end{document}